**Bright and fast voltage reporters across the visible spectrum via electrochromic FRET (eFRET)**

Peng Zou*[1], Yongxin Zhao*[2], Adam D. Douglass[3], Daniel R. Hochbaum[1], Daan Brinks[1], Christopher A. Werley[1], D. Jed Harrison[2], Robert E. Campbell[†2], Adam E. Cohen[§1,4]

[1]Department of Chemistry and Chemical Biology, Harvard University, Cambridge, MA 02138
[2]Department of Chemistry, University of Alberta, Edmonton, Alberta, Canada, T6G 2G2
[3]Department of Neurobiology and Anatomy, University of Utah, Salt Lake City, UT 84132
[4]Howard Hughes Medical Institute
* Equal contribution
† For correspondence regarding directed evolution of Arch: robert.e.campbell@ualberta.ca
§cohen@chemistry.harvard.edu

**Abstract**
We present a palette of brightly fluorescent genetically encoded voltage indicators (GEVIs) with excitation and emission peaks spanning the visible spectrum, sensitivities from 6 – 10% ∆F/F per 100 mV, and half-maximal response times from 1 – 7 ms. A fluorescent protein is fused to an Archaerhodopsin-derived voltage sensor. Voltage-induced shifts in the absorption spectrum of the rhodopsin lead to voltage-dependent nonradiative quenching of the appended fluorescent protein. Through a library screen, we identified linkers and fluorescent protein combinations which reported neuronal action potentials in cultured rat hippocampal neurons with a single-trial signal-to-noise ratio from 6.6 to 11.6 in a 1 kHz imaging bandwidth at modest illumination intensity. The freedom to choose a voltage indicator from an array of colors facilitates multicolor voltage imaging, as well as combination with other optical reporters and optogenetic actuators.

**Introduction**
Membrane voltage acts on all transmembrane proteins: the membrane electric field pulls on charged residues, shifting the free energy landscape for charge-displacing conformational transitions[1]. This bioelectric modulation is most famously observed in voltage-gated ion channels in neurons and cardiomyocytes, but voltage also affects the activity of G protein-coupled receptors (GPCRs) and some transmembrane enzymes[2]. Membrane voltage is dynamically regulated in bacteria[3], fungi[4], plants[5], and many cell types and sub-cellular organelles in the human body, and is disregulated in states of neuronal, cardiac, and metabolic diseases. Thus there is a need for fast, sensitive, bright, and spectrally tunable reporters of membrane voltage.

Recent progress in genetically encoded voltage indicators (GEVIs) has led to several classes of proteins which robustly report action potentials in cultured neurons. The fastest and most sensitive GEVIs are based on mutated microbial rhodopsins, in which membrane voltage modulates the endogenous fluorescence of the retinal chromophore[6]. Archaerhodopsin 3 (Arch) and its mutants are dim, though their high sensitivity and extreme photostability offset the low brightness from a signal-to-noise perspective. Other GEVIs are based on fusion of transmembrane voltage-sensing domains to fluorescent proteins. In some of these, voltage modulates the brightness of a single fluorescent fusion [7,8], while in others, voltage modulates the efficiency of Förster resonance energy transfer (FRET) between a pair of fluorescent fusions [9,10]. These fluorescent protein-based voltage sensors tend to have high brightness, but limited speed and sensitivity, and photobleaching can be a concern.

The availability of GEVIs spanning the visible spectrum is important when combining GEVIs with other GEVIs, other optical reporters, or optogenetic actuators. Having GEVIs of multiple colors enables multiplex voltage imaging when distinct structures cannot be spatially resolved. For instance, two colors of GEVIs would be required to image simultaneously the activity of excitatory and inhibitory neurons in intact tissue. Furthermore, green fluorescent protein (GFP)-based GEVIs cannot be paired with other GFP-based reporters such as GCaMP reporters of $Ca^{2+}$ (Ref. [11]), iGluSnFR reporter of glutamate[12], Perceval reporter of ATP[13], Clomeleon reporter of $Cl^-$ (Ref. [14]), and Pyronic reporter of pyruvate.[15] GFP-based reporters also experience severe optical crosstalk with all reported optogenetic actuators. Even the most red-shifted channelrhodopsin variants retain ~20% activation with blue light used for GFP excitation[16].

The excitation and emission wavelengths of a GEVI or other optical reporter also influence performance when imaging in intact tissue. The excitation and emission spectra of flavins, a major contributor to brain autofluorescence, overlap strongly with those of GFP [17, 18]. Consequently, a rule of thumb is that the signal-to-background ratio for GFP-based reporters is 10-fold lower in tissue than in cell culture. Redder reporters have significantly less background in brain tissue. Furthermore light scattering in tissue scales with wavelength $\lambda$ approximately as $\lambda^{-2.33}$, (Ref. [19]). Thus illumination at 640 nm propagates nearly 1.9-fold further into tissue than does illumination at 488 nm. While two-photon excitation of GFP-based reporters enables even greater depth penetration, this comes at the cost of requiring serial scanning, and a complex optical setup.

We sought a means to combine the high speed and sensitivity of Arch-based GEVIs with the brightness and spectral range of conventional fluorescent proteins. We used voltage-induced changes in the absorption spectrum of the retinal chromophore in Arch mutants to alter the degree of nonradiative quenching of a closely fused fluorescent protein (Fig. 1). Traditionally, FRET is used to measure the physical distance between a donor and acceptor. Here we use electrochromic changes in the acceptor to alter the overlap integral between the emission spectrum of the donor and the absorption spectrum of the acceptor. Thus we call this phenomenon electrochromic FRET (eFRET).

The voltage-dependent absorption spectrum of Arch has not been measured, but a detailed mechanistic study suggested that membrane depolarization increased the absorption in the yellow to red part of the spectrum.[20] We previously reported that in some microbial rhodopsins, conformation-induced absorption changes affected the degree of FRET from an appended fluorophore [21]. The present strategy is an outgrowth of that finding.

**Results**
Our electrochromic quencher is a mutant of Archaerhodopsin 3 (Arch), termed QuasAr2, which was shown to exhibit fast and sensitive changes in fluorescence in response to changes in membrane voltage [22]. The changes in fluorescence likely arose from changes in the absorption spectrum, so we reasoned that QuasAr2 would be an effective tool for voltage-dependent quenching of an appended fluorescent protein. Here we demonstrate fast and sensitive voltage-dependent changes in the fluorescence of fluorescent protein fusions spanning the visible spectrum: enhanced GFP (EGFP), yellow FP (Citrine), mOrange2, mRuby2, and mKate2. This work opens the possibility of multicolor voltage imaging with GEVIs.

The emission spectrum of mOrange2 [23] has a high degree of overlap with the absorption spectrum of Arch. To enable efficient FRET between the two proteins, it is critical to optimize the length of the linker connecting the two proteins, as rate of FRET is proportional to the reciprocal sixth power of the distance between the donor and acceptor chromophores. We examined the crystal structures of Arch-2 (a QuasAr2 homologue) (PDB ID 2EI4) [24] and mOrange (mOrange2 homologue) (PDB ID 2H5O) [25], and found that the last 15 C-terminal residues of Arch-2 and 15 N-terminal residues of mOrange are largely unstructured (Fig. 2A). We reasoned that those unstructured residues could be removed without disrupting the function of either protein. Therefore, we created a series of linker libraries by systematically truncating these resectable regions to shorten the distance between Arch and mOrange2 (Fig. 2B and Methods) and randomizing two residues at the junction.

We developed a hierarchical screening strategy to identify variants with optimal mOrange2 brightness, membrane trafficking, and voltage response (Fig. 2C and Methods). We used a customized vector, termed pcDuEx1.0, which is suitable for gene expression in both prokaryotic and eukaryotic cells. This vector circumvents the laborious cloning process typically required for switching between a primary bacteria-based screens and a secondary mammalian cell-based screen (Methods). Colonies of *E. coli* transformed with linker libraries in pcDuEx1.0 exhibited varied intensities of mOrange2 fluorescence. The colonies with brightest mOrange2 fluorescence were picked for secondary screening in mammalian cells. For the mammalian cell-based screen, a green GEVI, Arclight Q239 [7], was used as an internal reference for both membrane trafficking and voltage response (Fig. 2C-G and Methods). The sensitivity of eFRET variants generally increased as the linker was shortened. The most truncated variants ($\Delta 31$ and $\Delta 32$ libraries) had excellent voltage responses to membrane depolarizations, though this was counteracted by significantly decreased brightness of mOrange2, likely due either to poor folding efficiency caused by the aggressive truncation at the N-terminus, or to high basal quenching by the rhodopsin. In libraries with more than 20 residues truncated, combinations of Arg, Leu, Ala, and Gly were selected for at the two randomized residues at the junction connecting the two proteins. The variant from the $\Delta 24$ library with sequence 'Leu Arg' at the junction gave the best overall performance and was used henceforth. We added a trafficking sequence (TS) and endoplasmic reticulum export motif (ER2) to improve trafficking to the plasma membrane (Fig. 3B) [26].

We next characterized the optimized QuasAr2-mOrange2 construct in HEK cells and neurons. All measurements were performed at 23 °C. When expressed in HEK cells, the fusion protein efficiently trafficked to the plasma membrane (Fig. 3B) and exhibited bright mOrange2 fluorescence when excited with light at 532 nm (3 W/cm$^2$). To achieve similar fluorescence intensity via direct excitation of QuasAr2 at 640 nm required 70-fold greater illumination intensity (200 W/cm$^2$).

We used a patch pipette in the whole-cell configuration to vary the membrane voltage between -100 mV and +100 mV in a triangle wave at 0.5 Hz (Fig. 3C). Fluorescence traces were extracted from images of cells either by manually defining a region of interest around the cell, or by using an automated pixel-weighting algorithm[6]. Both approaches gave similar results. Fluorescence of mOrange2 showed nearly linear dependence on applied voltage, with a sensitivity of -9.5% $\Delta F/F$ per 100 mV, relative to fluorescence at -100 mV (Fig. 3D). The negative slope of dF/dV was consistent with increased absorbance of QuasAr2—and hence stronger quenching of mOrange2—at depolarizing voltages.

Fast fluorescence response to a change in voltage is essential for detection of action potentials in neurons. We applied a square wave of voltage (-70 mV to +30 mV, 5 Hz) and recorded the mOrange2 fluorescence at a frame rate of 2,000 frames/s (Fig. 3E). Fluorescence showed a biphasic step response, with a fast decay time constant of 2.4 ms (71% of the response) and a slow decay time constant of 22 ms (29% of the response) (Fig. 3F). The response was nearly symmetric between rising and falling edges (Table 1).

For comparison, the voltage response of QuasAr2 was also measured in the same construct. QuasAr2 fluorescence was more sensitive to voltage (ΔF/F = 90% per 100 mV, Fig. 3G) and had a somewhat faster response time (double-exponential fit to fluorescence decay: $\tau_1$ = 1.1 ms, 75%, $\tau_2$ = 14 ms, 25%, Fig. 3H). The discrepancy in response time between the QuasAr2 and mOrange2 may arise because, in addition to the fast fluorescence-determining transition, QuasAr2 may also undergo slower voltage-dependent transitions that change its absorption spectrum but that do not affect its fluorescence. These transitions could affect the quenching of the appended FP. Alternatively, voltage-induced conformational shifts in the QuasAr2 may exert stresses on the FP which directly modulate its quantum yield.

We also compared the performance of the eFRET GEVI to Arclight Q239, a GFP-based GEVI.[7] Arclight showed voltage sensitivity of -32% ΔF/F per 100 mV (Fig. 3F), consistent with previous reports and 3.3-fold more sensitive than the mOrange2 eFRET GEVI. The step response of Arclight followed a single exponential rise with a time constant of 100 ms (for a voltage step from +30 mV to -70 mV) and a bi-exponential decay with time constants of 16 ms (52%) and 140 ms (48%). These numbers are slower than in the original report on Arclight[7], likely due to the difference in measurement temperature: 23 °C in our measurements vs. 33-35 °C in the prior report. As measured by the time to achieve half-maximal step response, the mOrange2 eFRET GEVI was 19-fold faster than Arclight on the fluorescence rise, and 11-fold faster on the decay, at room temperature (Table 1).

Arch absorbs strongly from 500 – 640 nm, so we reasoned that eFRET might work for fluorescent proteins spanning the visible spectrum. We created a palette of eFRET constructs by fusing ECFP[27], EGFP[27], Citrine[28], mRuby2[10], and mKate2[29] to the C-terminus of QuasAr2, in similar fashion to the QuasAr2-mOrange2 fusion. We tested the voltage sensitivity and step response as above. Three of the new constructs, QuasAr2-EGFP, QuasAr2-Citrine and QuasAr2-mRuby2, showed sensitivity and speed comparable to the mOrange2 construct (Fig. 4 and Table 1). The bluest and reddest constructs, containing ECFP and mKate2, exhibited poor voltage sensitivity, which we attribute to poor spectral overlap with QuasAr2 absorption.

We further characterized the most sensitive eFRET voltage sensors via transient transfection in cultured rat hippocampal neurons. Constructs exhibited good trafficking to the plasma membrane (Fig. 5). Injection of current pulses via a patch pipette (500 – 600 pA, 5 – 10 ms, 5 Hz) induced trains of action potentials, which induced downward fluorescence transients of 12% (Citrine), 12% (mOrange2), and 7% (mRuby2). Arclight also reported single-trial action potentials when used at 34 °C, with a downward transient of 3%. Despite the greater steady-state voltage sensitivity of Arclight than the eFRET GEVIs, the slow response of Arclight acted as a low-pass filter on millisecond-timescale action potentials. eFRET constructs also reported spontaneous activity in the absence of a patch pipette, clearly resolving subthreshold dynamics and single action potentials (Fig. 6).

The signal-to-noise ratio (SNR) of fluorescence detection, defined as the ratio of the peak amplitude to standard deviation of fluorescence at the baseline, was 6.6 (Citrine), 11.6 (mOrange2), and 10.5 (mRuby2) (1 kHz frame rate, illumination at 3 $W/cm^2$, 23 °C for all eFRET GEVIs). Parallel measurements on Arclight had a single-trial SNR of 6.2 and drastically distorted the AP waveform (1 kHz frame rate, illumination at 10 $W/cm^2$, 34 °C). Previous measurements with QuasAr2 reported a SNR of 30-70 at the same bandwidth but with 800 $W/cm^2$ illumination [22].

**Discussion**
We developed a palette of fluorescent GEVIs spanning the visible spectrum: absorption and emission spectra were set by the eFRET donor, which was selected from EGFP, Citrine, mOrange2, mRuby2, or

mKate2. The sensors based on Citrine, mOrange2, and mRuby2 had sufficient sensitivity and speed to resolve single-trial action potentials in cultured neurons, and each was spectrally distinguishable from the others. While the GEVI Arclight has larger voltage sensitivity, the greater speed of the eFRET GEVIs led to higher SNR in detection of single action potentials from cultured neurons. The illumination intensities needed to image eFRET GEVIs are compatible with LED illumination, greatly decreasing the cost and complexity of instrumentation compared to that needed to image retinal fluorescence directly. The fluorescent proteins used in eFRET GEVIs have strong 2-photon cross sections, and thus these proteins may facilitate 2-photon voltage imaging in tissue.

Single-color voltage imaging has severe limits on the types of signals it can probe. Wide-field mapping of field potentials in brain slice or *in vivo* cannot report the simultaneous dynamics of multiple defined cell types. Thus one cannot probe the relative contributions of excitatory and inhibitory neurons; or how the signals from glia relate to nearby neural activity. Recent attempts at voltage imaging *in vivo* used a GEVI under a ubiquitous CAG promoter, and thus reported signals averaged over all cells within the imaged brain region [30]. Similarly, bath-applied voltage-sensitive dyes could only report activity of brain regions, not of cell types [31].

With simultaneous multicolor voltage imaging one could combine genetic targeting and multicolor imaging to record voltage simultaneously from distinct structures that might otherwise be difficult to distinguish. One could map activity simultaneously in excitatory and inhibitory neuronal populations; or in sub-populations distinguished by even finer genetic differences. With simultaneous multicolor voltage imaging one could probe the underlying mechanisms of excitation-inhibition balance, in health and in disease. A variation of this scheme is to express in each cell a randomly selected subset of the eFRET GEVIs, following the methodology used to generate the Brainbow palette [32]. By this approach, neighboring cells in intact tissue would each report their electrical activity in a unique hue, enabling simultaneous structural and functional mapping. The broad palette of eFRET GEVIs also facilitates combination with other GFP-based fluorescent reporters or a wide variety of optogenetic actuators.

**Acknowledgments**
We thank V. Nathan and N. Smedemark-Margulies for technical assistance, and V. Venkatachalam, S. Farhi, and J. Kralj for helpful discussions. This work was supported by the Harvard Center for Brain Science, PECASE award N00014-11-1-0549, US National Institutes of Health grants 1-R01-EB012498-01 and New Innovator grant 1-DP2-OD007428, the Natural Sciences and Engineering Research Council of Canada (Discovery grants to R.E.C.), the Canadian Institutes of Health Research (R.E.C.), and graduate scholarships from the University of Alberta and Alberta Innovates (Y.Z.). R.E.C. holds a Tier II Canada Research Chair.

**Author contributions**
PZ, YZ, DRH, ADD, CAW, DB conducted experiments and analyzed data. REC supervised the screen for linker length optimization. AEC proposed eFRET for voltage sensing. PZ, YZ, REC, and AEC wrote the paper.

| Fluorophore | Ex. max (nm) | Em. Max (nm) | Extinction coeff. ($M^{-1}\ cm^{-1}$) | Quantum yield | $-\Delta F/F$ per 100 mV (%) | Rising edge $\tau_{1/2}$ (ms) | Falling edge $\tau_{1/2}$ (ms) |
|---|---|---|---|---|---|---|---|
| ECFP | 433 | 475 | 32500 | 0.4 | 5.2 | n.d. | n.d. |
| EGFP | 488 | 507 | 56000 | 0.6 | 8.1 | 6.0 | 7.0 |
| Citrine | 516 | 529 | 77000 | 0.76 | 10.1 | 1.9 | 0.6 |
| mOrange2 | 549 | 565 | 58000 | 0.60 | 9.5 | 3.5 | 2.6 |
| mRuby2 | 559 | 600 | 113000 | 0.38 | 9.4 | 3.3 | 7.0 |
| mKate2 | 588 | 633 | 62500 | 0.40 | 6.1 | 1.1 | 1.6 |
| | | | | | | | |
| Arclight Q239 | | | | | 32 | 68 | 28 |

**Table 1.** Spectral properties of fluorescent proteins and voltage sensing properties of corresponding eFRET GEVIs. Data in shaded columns is from Ref. [33]. Response times were characterized by $\tau_{1/2}$, the time to reach 50% of the transition in a step of membrane voltage. Rising edge indicates fluorescence increase (voltage step from +30 mV to -70 mV). All voltage-dependent measurements were performed at 23 °C. Photophysical properties of Arclight Q239 have not been characterized, though the protein is spectrally similar to EGFP.

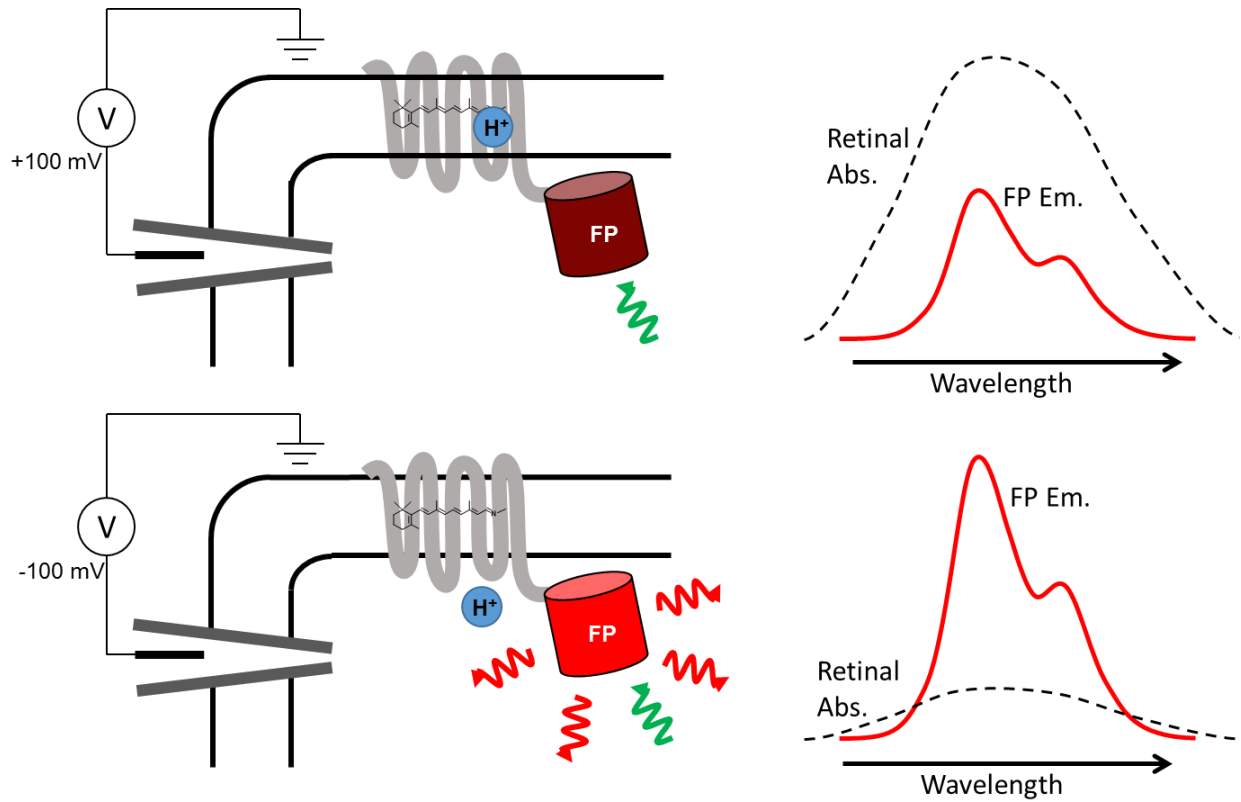

**Figure 1. Proposed mechanism of voltage-dependent fluorescence in eFRET-based GEVIs.** Left: cartoons showing the protein in a cell plasma membrane. Voltage controls the protonation of the Schiff base joining the retinal to the protein scaffold. The protonated state absorbs strongly with a peak near 600 nm, the deprotonated state does not. Right: cartoon of voltage-dependent absorption spectrum of the microbial rhodopsin (dashed line), and emission spectrum of the attached fluorescent protein. Top: at depolarizing (positive) membrane voltage, the microbial rhodopsin absorbs strongly and quenches the fluorescence of the fluorescent protein. Bottom: at hyperpolarizing (negative) membrane voltage, the microbial rhodopsin absorbs weakly, so the fluorescent protein emits strongly.

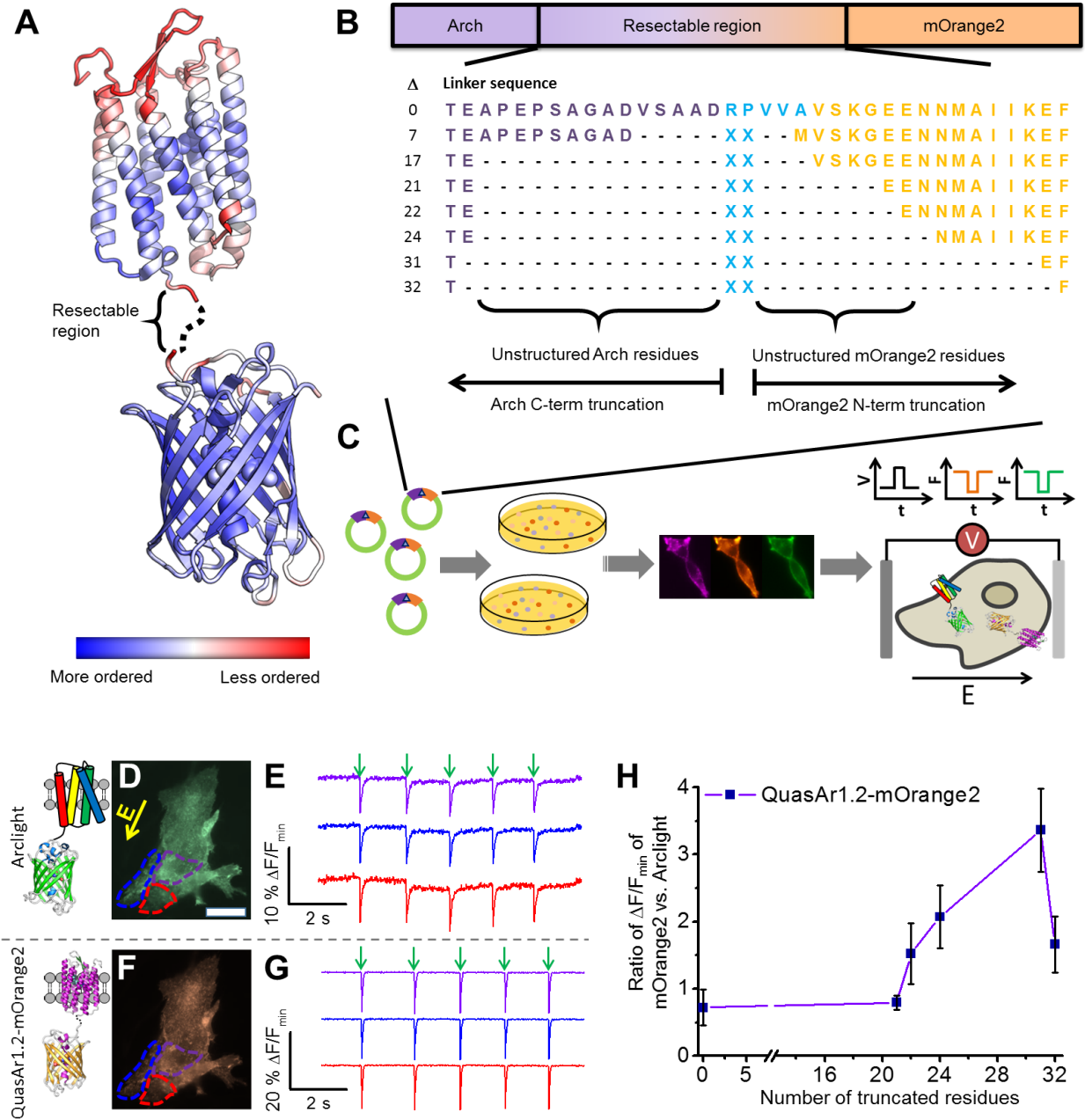

**Figure 2. Directed evolution of an eFRET GEVI.** (A) A model of Arch-mOrange2 constructs, represented by the crystal structures of Arch-2 (PDB ID 2EI4) [24] and mOrange (PDB ID 2H5O) [25]. The colorscale represents the degree of order in the crystal structure, as reported by B-factor. Less ordered regions are presumed more dispensable. (B) Design of linker truncation libraries. QuasAr1.2, the best available variant at the time of these experiments, was used as the electrochromic quencher. (C) Hierarchical screen of truncated linker libraries for eFRET GEVIs. Constructs were first screened in *E. coli* for mOrange2 brightness and then screened for membrane trafficking in HEK cells. Voltage-sensitivity was then tested via field stimulation in HeLa cells co-expressing the eFRET GEVI, Arclight (as an internal control), and Kir2.1 (to lower the resting voltage to -60 mV [34]). (D) Arclight fluorescence of three HeLa cells. Yellow arrow indicates direction of the applied electric field. Scale bar: 25 μm. (E) Fluorescence

intensity traces of the three regions shown in (D).  (F) mOrange2 fluorescence from the same cells shown in (D). (G) Response of mOrange2 fluorescence during stimulation as in (E). (H) Effect of linker length on voltage sensitivity in QuasAr1.2-mOrange2.  For each linker length the most sensitive eFRET GEVI construct was compared to Arclight measured in the same cells.  Error bars represent standard deviation over *n* = 15-20 cells in the ratio ΔF/F$_{min}$.

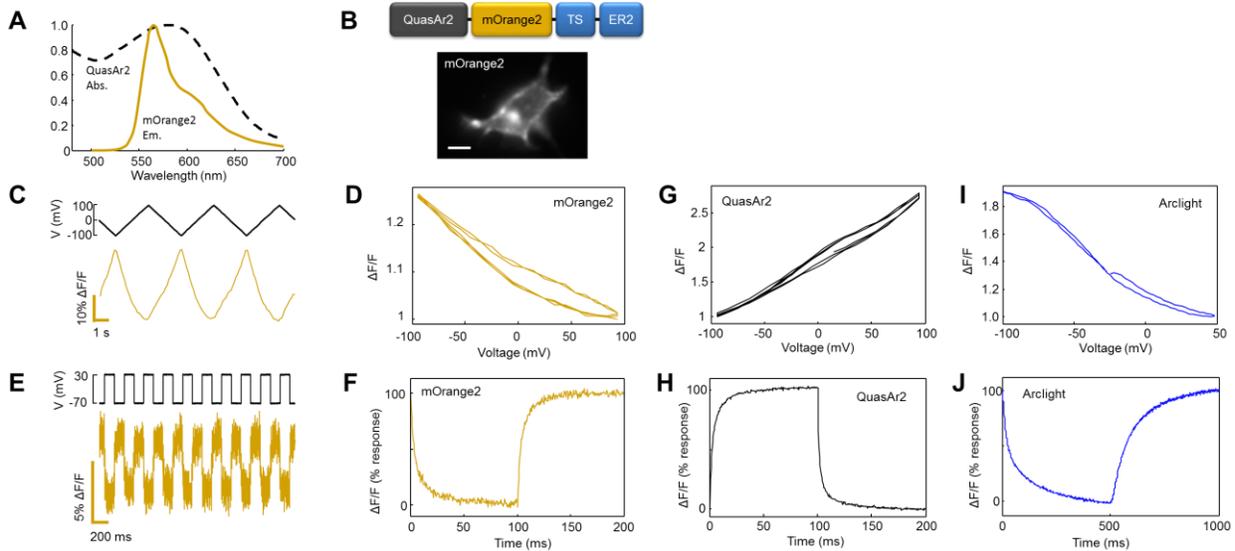

**Figure 3. Characterization of a red-shifted GEVI, QuasAr2-mOrange2.**  A) Absorption spectrum of QuasAr2 (black dashed) overlaid on emission spectrum of mOrange2 (orange solid).  B) Top: eFRET GEVI domain structure.  Bottom: Image of a HEK293 cell expressing the mOrange2 eFRET GEVI, imaged via mOrange2 fluorescence.  Membrane-localized protein is visible in the periphery of the cell, as well as untrafficked protein surrounding the nucleus and in internal puncta.  Scale bar 10 μm.  C) Fluorescence response to a triangle wave in membrane potential.  D) mOrange2 fluorescence as a function of membrane voltage.  Many eFRET-based GEVIs show a small amount of hysteresis.  E) Fluorescence response to a square wave in membrane potential, recorded at a frame rate of 1 kHz.  F) Mean mOrange2 step response.  G) QuasAr2 fluorescence as a function of membrane voltage, recorded in a QuasAr2-mOrange2 fusion.  H) Mean QuasAr2 step response, recorded in a QuasAr2-mOrange2 fusion. I) Arclight fluorescence as a function of membrane voltage.  J) Mean Arclight step response.  Note the different time axis from panels F and H.

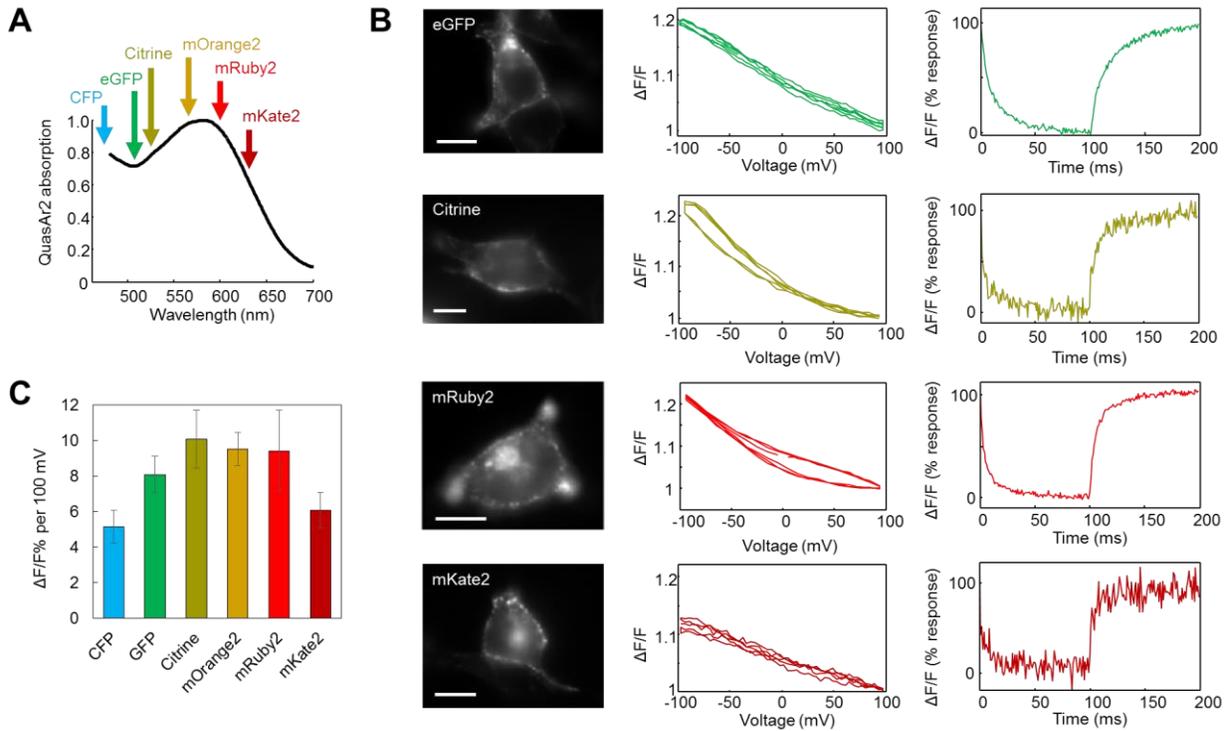

**Figure 4. eFRET GEVIs across the visible spectrum.** A) Emission maxima of the donor fluorescent proteins overlaid on the absorption spectrum recorded from purified QuasAr2. B) Representative single-cell images, plots of F vs. V, and step responses for four of the eFRET GEVIs tested. In all cases, the GEVI was expressed in HEK293 cells, and the voltage-sensing responses were characterized by manual patch-clamp electrophysiology. Scale bar 10 µm. C) Voltage-sensitivity of eFRET GEVIs approximately corresponds to the degree of spectral overlap between the emission of the GEVI and the absorption of QuasAr2.

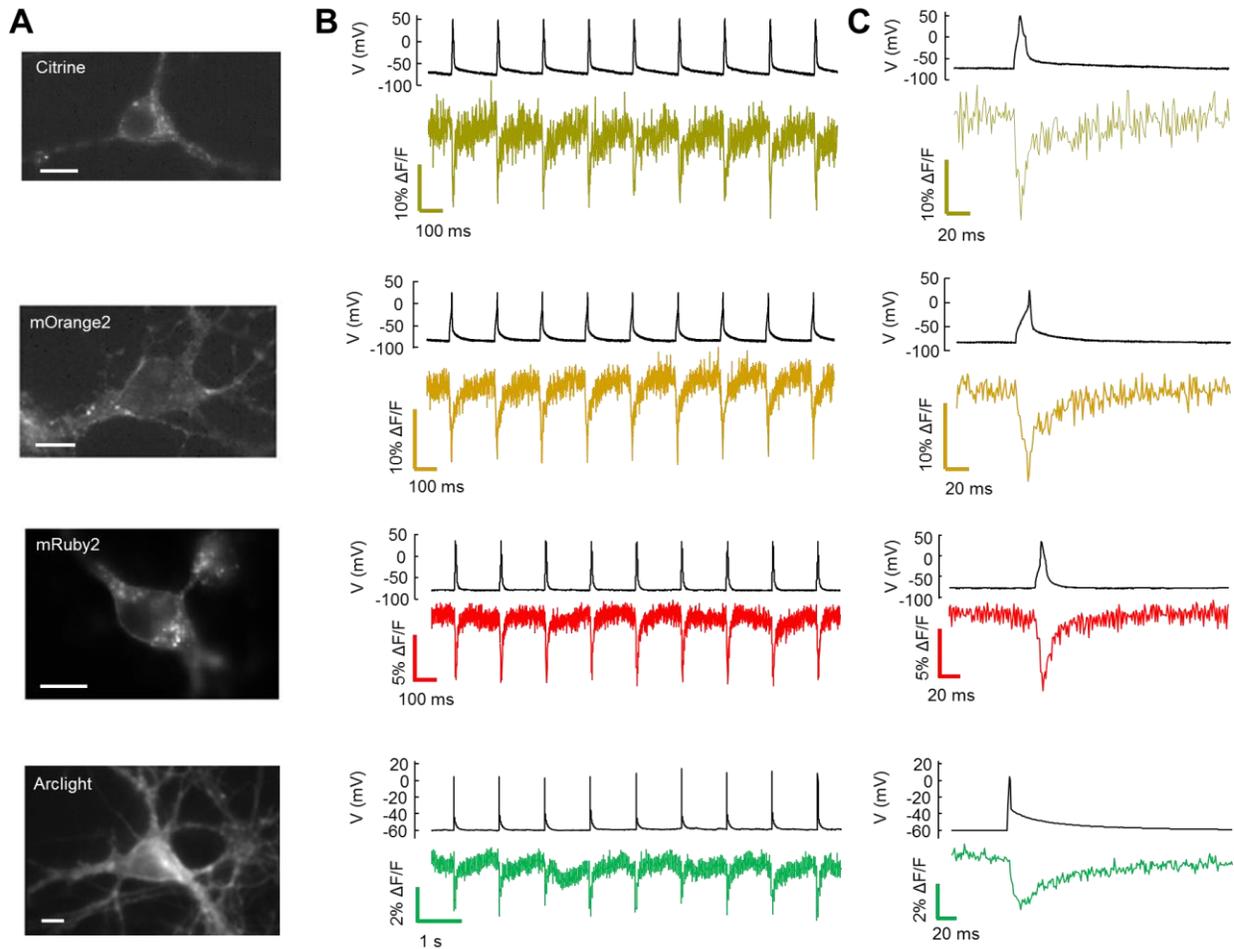

**Figure 5. Single-trial recording of neuronal action potentials with eFRET GEVIs and Arclight.** The GEVI construct was expressed in primary rat hippocampal neurons under a CamKIIα promoter. Action potentials were induced by current injections through a patch pipette. A) Images of neurons expressing the indicated GEVI. Scale bars 10 μm. B) Simultaneous patch clamp (top) and fluorescence (bottom) recordings of action potential waveforms. C) Close-up showing the electrically and optically reported action potential waveforms.

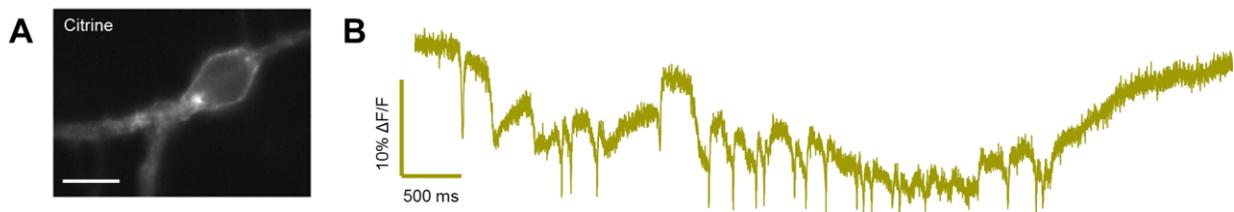

**Figure 6. Optical recording of spontaneous electrical activity.** In a neuron synaptically coupled to a spontaneously active circuit, the Citrine-based eFRET GEVI clearly revealed action potentials and sub-threshold depolarizations. A) Image of the neuron. Scale bar 10 μm. B) Optical recording from the cell.

**Materials and Methods**
**1. Materials**
Synthetic DNA oligonucleotides were purchased from Integrated DNA Technologies. *PfuUltraII* polymerase (Agilent Technologies) or Phusion Polymerase (New England Biolabs) were used for high fidelity PCR amplifications in the buffer supplied by the respective manufacturer. PCR products and products of restriction digests were purified either with PCR clean up kit (QIAGEN) or using preparative agarose gel electrophoresis followed by DNA isolation using the Zymoclean gel DNA recovery kit (Zymo Research). Restriction endonucleases were purchased from New England Biolabs and used according to the manufacturer's instructions. Ligations were performed using T4 DNA ligase (Invitrogen) or Gibson Assembly (New England Biolabs). Small-scale isolation of plasmid DNA was performed by plasmid miniprep kit (QIAGEN). The cDNA sequences for all fusion constructs were confirmed by sequencing (Genewiz). Site-directed mutagenesis was performed with QuikChange kit (Agilent Technologies).

**2. Construction of linker libraries**
A vector for expression in prokaryotic and eukaryotic systems was constructed based on mammalian expression vector pcDNA3.1 (+). To facilitate prokaryotic expression, a customized constitutive promoter was introduced (ttgctttgtgagcggataacaattataatagattca) based on the phage early T5 promoter for prokaryotic transcription [35]. An *E. coli* ribosome binding site (aggaggaa) for prokaryotic translation was also introduced via Quikchange reactions. The resultant vector, designated pcDuEx1.0, exhibits moderate expression of Arch-mOrange2 fusions in *E. coli* cells and very similar expression levels in HeLa cells compared to its predecessor. We used pcDuEx1.0 as the vector for screening of linker libraries.

We focused on mOrange2 fusions for determining optimized linker length for FRET efficiency. We chose QuasAr1.2 (Arch D95N/D106H, the best available at the time of the screen) as the Arch body. For construction of QuasAr1.2-mOrange2 $\Delta$0, we fused QuasAr1.2 to the N-terminus of mOrange2 fusion via a 5 residue linker (RPVVA) using overlap PCR. The DNA was PCR amplified with the flanking restriction sites BamHI and XbaI, followed by double digestion and ligation into pcDuEx1.0 linearized by digestion with the same two enzymes.

From the template of QuasAr1.2-mOrange2 $\Delta$0, we constructed various linker libraries by systematic truncation of the connecting region between QuasAr1.2 and mOrange2 using Quikchange reactions (Figure 2). For each library, 2 randomized amino acid residues (nucleotide sequence NNKNNK, where N = A, G, C, T and K = G, T) were placed between the C-terminus of Arch and the N-terminus of mOrange2, to generate 400 amino acid combinations (1024 nucleotide combinations). The libraries were cloned into pcDuEx1.0 via standard procedure of restriction enzyme digestion and ligation.

**3. Hierarchical screen of linker libraries**
*E. coli* colonies expressing the linker library exhibited various intensities of mOrange2 fluorescence. For each library, the 24 variants with brightest mOrange2 fluorescence were picked and grown in overnight culture at 37 °C. The plasmid DNA of each variant was prepared from overnight culture using standard mini-prep procedure. For each variant, the plasma membrane trafficking was examined in HEK cells co-expressing Arclight that served as internal reference. The 5 variants with the best membrane trafficking were then expressed in HeLa cells for testing of voltage sensitivity.

HeLa cells were grown to 40-60% confluence on home-made 35 mm glass bottom dishes or 24-well glass bottom plates, and transfected with 1 µg of plasmid DNA and 2 µL Turbofect (Thermo Scientific) according to the manufacturer's instructions. HeLa cells were transfected with plasmids encoding the QuasAr1.2-mOrange2 variant, Arclight Q239 (Addgene: 36856) and Kir2.1 (Addgene: 32641) in equal weight ratio. Expression of Kir2.1 in HeLa cells maintained the resting potential close to -60 mV, appropriate for a neuronal voltage indicator [34]. After 3 h incubation, the media was exchanged to DMEM with 10% fetal bovine serum and the cells were incubated for an additional 24 h at 37 °C in a $CO_2$ incubator. Immediately prior to imaging, cells were washed twice with Hanks balanced salt solution (HBSS) and then 1 mL of 20 mM HEPES buffered HBSS was added. No retinal was added to the buffer-- loading of retinal into the rhodopsin was presumed to occur from endogenous retinal.

Cell imaging was performed with an inverted Eclipse Ti-E (Nikon) equipped with a Photometrics QuantEM 512SC camera and a 150 W mercury-xenon lamp (Hamamatsu). A home-made parallel platinum electrode pair with a separation distance of 0.5 cm was mounted in a custom plastic support and was placed in the imaging dish or well. The waveforms of voltage pulses were generated by a pulse generator PG 58A (Gould Advance Ltd.) and amplified by an Agilent 6824A 40V/25A DC Power Supply (Hewlett Packard). The typical waveform had square wave pulses lasting 20 ms with pulse field strength ranging from 50 - 60 V/cm. The mOrange2 fluorescence was imaged at 100 Hz frame rate in 4×4 binning mode for 10 s using the following filter set: 545/30 nm (excitation), 620/60 nm (emission), and 565 nm (dichroic). For imaging Arclight, the filter set was: 480/40 nm (excitation), 535/40 nm (emission), and 505 nm (dichroic).

The raw fluorescence traces of both mOrange2 and Arclight were extracted from identical regions of interest in cells expressing both constructs, and exported into a Microsoft Excel spreadsheet using the microscope software NIS-Elements Advanced Research (Nikon). Background subtraction, photobleaching corrections, calculations of average $\Delta F/F_{min}$, and calculation of signal-to-noise ratios (SNR) were performed automatically in Excel. The average $\Delta F/F_{min}$ and SNR of mOrange2 signals was compared to those of Arclight signals from the same cells, and the ratios of $\Delta F/F_{min}$ of mOrange2 vs. Arclight were reported. At least 20 cells co-expressing mOrange2 fusion and Arclight were analyzed for each variant. The best variant with maximum mean ratio in each library was determined and sequenced.

### 4. Expression vectors for HEK cells and neurons
We chose lentivirus vector FCK-Arch-EGFP (Addgene: 22217) as the backbone for all eFRET constructs. This vector features a *CaMKIIα* promoter and a Woodchuck Hepatitis Virus Posttranscriptional Regulatory Element (WPRE) after the 3' end of the open reading frame. To enhance membrane trafficking of fusion proteins, we added a trafficking signal (TS) and ER export signal peptide sequence (FCYENEV), derived from the inward rectifier potassium channel Kir2.1, as previously described [26]. QuasAr2-FP fusion constructs were made by Gibson Assembly: the vector was linearized by double digestion with BamHI and BsrGI, and QuasAr2 and fluorescent protein cDNA segments were generated by PCR amplification.

The linker configuration of all eFRET fusion proteins was constructed based on the optimized linker sequence found by the linker screening. In all eFRET fusion proteins, the N-terminal 2 amino acids of QuasAr2 were changed from DP to VS to facilitate expression, and the C-terminal 14 amino acids of QuasAr2 (APEPSAGADVSAAD) were replaced with a 2-amino acid linker, LR (based on the optimized linker sequence found in linker library Δ24), to shorten the distance between QuasAr2 and fluorescent protein. The N-terminal amino acids in the fluorescent fusions were also truncated, as listed in Table 2.

| Fluorescent protein | N-terminal truncations |
|---|---|
| ECFP | MSKGEEL |
| EGFP | M |

| Citrine  | MSKGEEL  |
|----------|----------|
| mOrange2 | MVSKGEEN |
| mRuby2   | MVSKGEEL |
| mKate2   | M        |

**Table 2.** Truncation of N-terminal amino acids in fluorescent proteins

## 5. Simultaneous electrophysiology and fluorescence in HEK cells

HEK293 cells were cultured and transfected following standard protocols. Briefly, HEK293 cells were grown at 37 °C, 5% $CO_2$, in DMEM supplemented with 10% FBS and penicillin-streptomycin. Plasmids were transfected using TransIT-293 reagent (Mirus Bio LLC) following the manufacturer's instructions. 24 hours post-transfection, cells were re-plated onto glass-bottom dishes (MatTek) at a density of ~10,000 cells/$cm^2$. Cells were assayed 40-60 hours post-transfection.

Cells were supplemented with retinal by pre-incubating with 5 μM retinal in growth medium (diluted from 40 mM stock solution in DMSO) in the incubator for 0.5 - 1 hour immediately prior to imaging. All imaging and electrophysiology were performed in Tyrode's buffer (containing 125 mM NaCl, 2.5 mM KCl, 3 mM $CaCl_2$, 1 mM $MgCl_2$, 10 mM HEPES, 30 mM glucose, at pH 7.3, and adjusted to 305-310 mOsm with sucrose). A gap junction blocker, 2-aminoethoxydiphenyl borate (50 μM, Sigma), was added to eliminate electrical coupling between cells.

Filamented glass micropipettes (WPI) were pulled to a tip resistance of 5–10 MΩ, and filled with internal solution containing 125 mM potassium gluconate, 8 mM NaCl, 0.6 mM $MgCl_2$, 0.1 mM $CaCl_2$, 1 mM EGTA, 10 mM HEPES, 4 mM Mg-ATP, 0.4 mM Na-GTP (pH 7.3); adjusted to 295 mOsm with sucrose. Pipettes were positioned with a Sutter MP285 manipulator. Whole-cell, voltage and current clamp recordings were acquired using a patch clamp amplifier (A-M Systems, Model 2400), filtered at 5 kHz with the internal filter and digitized with a National Instruments PCIE-6323 acquisition board at 10 kHz.

Simultaneous whole-cell patch clamp recordings and fluorescence recordings were acquired on a home-built, inverted epifluorescence microscope, described below in the section "Imaging Apparatus".

## 6. Neuronal culture, gene delivery and electrophysiology

All procedures involving animals were in accordance with the National Institutes of Health Guide for the care and use of laboratory animals and were approved by the Institutional Animal Care and Use Committee (IACUC) at the institution at which they were carried out.

Hippocampal neurons from P0 rat pups were dissected and cultured in neurobasal-based medium (NBActiv4, Brainbits llc.) at a density of 40,000 $cm^{-2}$ on glass-bottom dishes (MatTek) pre-coated with poly-d-lysine (Sigma P7205) and matrigel (BD biosciences 356234). At 3 days *in vitro* (DIV), cytarabine was added to the neuronal culture medium at a final concentration of 2 μM to inhibit glial growth [36].

Neurons were transfected on DIV 7 with the eFRET plasmid using Lipofectamine 2000 transfection reagent (Life Technologies). Procedures followed manufacturer's instructions but reduced the amount of reagent by 50-80% to avoid toxicity.

Measurements were performed on primary cultures at DIV 10 - 20. Experiments were conducted in Tyrode's solution containing 125 mM NaCl, 2.5 mM KCl, 3 mM $CaCl_2$, 1 mM $MgCl_2$, 10 mM HEPES, 30 mM glucose (pH 7.3) and adjusted to 305–310 mOsm with sucrose. Immediately prior to imaging, neurons were incubated with 5 µM all-*trans* retinal in the culture medium for 30 minutes and then washed with Tyrode's solution. Experiments were performed at 23 ºC under ambient atmosphere.

### 7. Imaging apparatus

Experiments were conducted on a home-built inverted fluorescence microscope equipped with 488nm, 532nm, 561nm, 594nm, and 640nm laser lines and a scientific CMOS camera (Hamamatsu ORCA-Flash 4.0). The power and manufacturer of laser lines are summarized in Table 3. Illumination from lasers were pre-combined using dichroic mirrors, sent through an acousto-optic tunable filter (AOTF; Gooch and Housego 48058-2.5-.55-5W) for intensity modulation control, and then expanded and focused onto the back-focal plane of a 60× water immersion objective, numerical aperture 1.20 (Olympus UIS2 UPlanSApo 60x/1.20 W). Imaging of fluorescent proteins was performed at illumination intensities of 2 – 4 W/cm$^2$. Imaging of QuasAr2 direct fluorescence was performed at an illumination intensity of 400 W/cm$^2$. Table 4 summarizes the laser lines, dichroic mirrors and emission filters used for fluorescence imaging. For fast data acquisition, a small field of view around the cell of interest was chosen at the center of the camera to achieve a frame rate of 1,000 frames per second.

### 8. Data analysis

Data analysis was done with homemade software written in MATLAB. Fluorescence intensities from raw movies were extracted using a maximum likelihood pixel weighting algorithm described in Ref. [6]. Briefly, the fluorescence at each pixel was correlated with the applied voltage. Pixels that showed stronger correlation to the mean were preferentially weighted. This algorithm automatically found the pixels carrying the most information, and de-emphasized background pixels. Alternatively, a region of interest (ROI) comprising the cell body was defined by the user, and fluorescence intensity was calculated from the unweighted mean of pixel values within the ROI. With the improved trafficking of the Arch mutants resulting from the TS and ER motifs, the ROI approach gave similar results as the maximum likelihood pixel weighting algorithm.

For eFRET speed analysis, the time constants for the step response were calculated by fitting a double exponential to the rising and decaying portions of the fluorescence traces.
All error ranges represent standard error of the mean.

| Wavelength | Laser name | Maximum Power |
|---|---|---|
| 447 nm | Dragon Lasers 447M100 | 100 mW |
| 488 nm | Coherent Obis 488-50 | 50 mW |
| 532 nm | Dragon Lasers MGL-III-532 | 50 mW |
| 561 nm | Cobalt Jive 0561-04-0-0150-300 | 150 mW |
| 594 nm | Cobalt Mambo 0594-04-01-0100-300 | 100 mW |
| 640 nm | Coherent Obis 637-140 LX | 140 mW |

**Table 3.** Lasers used in home-built imaging apparatus

| Fluorophore | Excitation | Dichroic | Emission |
|---|---|---|---|
| ECFP | 447 nm | Semrock Di02-R442-25x36 | 480/40 |
| EGFP | 488 nm | Chroma zt505-515+650NIR Tpc | 525/30 |
| Citrine | 488 nm | Semrock Di01-R405/488/594-25x36 | 525/30 |
| mOrange2 | 532 nm | Chroma zt532/640rpc, | 575/50, or Semrock |

|  |  | or zt532/635rpc | Di01-R405/488/532/635 |
| --- | --- | --- | --- |
| mRuby2 | 561 nm | Chroma zt488/561/635rpc | 600/37 |
| mKate2 | 594 nm | Semrock Di01-R405/488/594-25x36 | 620LP |
| Arch | 640 nm | Chroma zt505-515+650NIR Tpc, or zt532/640rpc, or zt488/561/635rpc | 708/75 |

**Table 4.** Filter sets used in fluorescence imaging